\documentclass[aps,prd,preprint,preprintnumbers,nofootinbib,superscriptaddress]{revtex4}
\usepackage{ae}
\usepackage{bm} 
\usepackage{color}
\usepackage{amssymb}
\usepackage{amsmath}
\usepackage{amsfonts}
\usepackage{graphicx}
\usepackage{slashed}
\usepackage{setspace}
\usepackage{subfigure}
\usepackage{ulem}
\usepackage{bbold}
\usepackage{upgreek}

\newcommand{\dns}[1]{} 
\newcommand{\Eqref}[1]{Eq.~\eqref{#1}}

\allowdisplaybreaks

\begin{document}

\date{\today}

\title{Vacuum birefringence at the Gamma Factory}

\author{Felix Karbstein}
\email{felix.karbstein@uni-jena.de}
\affiliation{Helmholtz-Institut Jena, Fr\"obelstieg 3, 07743 Jena, Germany}
\affiliation{GSI Helmholtzzentrum f\"ur Schwerionenforschung, Planckstra\ss e 1, 64291 Darmstadt, Germany}
\affiliation{Theoretisch-Physikalisches Institut, Abbe Center of Photonics, \\ Friedrich-Schiller-Universit\"at Jena, Max-Wien-Platz 1, 07743 Jena, Germany}

\begin{abstract}
We explore the perspectives of studying vacuum birefringence at the Gamma Factory. To this end, we assess in detail the parameter regime which can be reliably analyzed resorting to the leading contribution to the Heisenberg-Euler effective Lagrangian.
We explicitly show that -- contrarily to naive expectations -- this approach allows for the accurate theoretical study of quantum vacuum signatures up to fairly large photon energies.
The big advantage of this parameter regime is the possibility of studying the phenomenon in experimentally realistic, manifestly inhomogeneous pump and probe field configurations.
Thereafter, we focus on two specific scenarios giving rise to a vacuum birefringence effect for traversing gamma probe photons.
In the first scenario the birefringence phenomenon is induced by a quasi-constant static magnetic field.
In the second case it is driven by a counter-propagating high-intensity laser field.
\end{abstract}

\maketitle


\section{Introduction}

Maxwell's classical theory of electrodynamics allows for the accurate description of the physics of macroscopic electromagnetic fields and light propagation.
One of its cornerstones is the superposition principle of electromagnetic waves in the vacuum.
The latter implies that in vacuum light rays pass through each other without any interaction and thus do not change their properties.
This celebrated classical principle does, however, no longer strictly hold when resorting to a modern theoretical physics perspective in which the microscopic physics is governed by a relativistic quantum field theory (QFT).

A simplified intuitive explanation goes as follows.
As opposed to the classical notion of the vacuum, the {\it quantum vacuum} can no longer be considered as an empty and inert ground state of the world, but amounts to a highly non-trivial quantum state which encodes information about the full particle content of the underlying fundamental quantum theory.
This comes about as follows: QFT distinguishes between {\it real} and {\it virtual} manifestations of particles, both of which are characterized by the same defining properties, namely mass, charge and spin.
The real variant fulfills a relativistic energy-momentum relation (``on-shell condition'') and may be considered as the quantum version of a classical particle. 
It exists as an asymptotic state, and can -- at least in principle -- be prepared and measured directly in experiment.
In contrast, virtual particles violate the relativistic energy-momentum relation, i.e., are manifestly off-shell, and have no classical analogue.
They only exist as internal lines in the Feynman diagrams mediating the microscopic particle physics interaction processes, but not as external lines representing in- and outgoing asymptotic states, inhibiting their direct preparation or measurement. At the same time, conserved quantities, such as energy, momentum, charge, spin, etc., are manifestly conserved at each interaction, independently if this involves just real, just virtual or both variants of particles. In determining the transition amplitude for a physical, gauge invariant process characterized by given in- and out-states, one has to sum over all possible virtual-particle exchanges mediating between these in- and out states.

Adopting this picture, the quantum vacuum is characterized by the omnipresence of virtual particle-antiparticle fluctuations, which can be probed by real particles or fields.
As these fluctuations involve all particle degrees of freedom of the underlying quantum theory, the quantum vacuum even constitutes a portal to physics beyond the Standard Model (SM) of particle physics.
In the present study we limit our discussion to quantum vacuum physics within the SM. A quantitative understanding of the effects predicted within the SM is absolutely indispensable for performing reliable studies of the impact of beyond the SM particles on quantum vacuum signatures.

As has been realized already in the 1930s by Heisenberg and Euler \cite{Heisenberg:1935qt} the fluctuations of electrons and positrons can give rise to effective nonlinear interactions of prescribed electromagnetic fields in the vacuum.
This directly results in light-by-light scattering phenomena  and violations of the classical superposition principle \cite{Euler:1935zz,Karplus:1950zza,Karplus:1950zz,Costantini:1971cj,DeTollis:1965vna}.

The fundamental parameters of quantum electrodynamics (QED) imprinted on the quantum vacuum are the electron mass $m_e\simeq511\,{\rm keV}/c^2$ and charge $e$.
Upon combination with the speed of light $c$ and Planck's constant $\hbar$, these parameters can be converted into reference field strengths, namely the so-called {\it critical} electric $E_{\rm cr}=m_e^2c^3/(e\hbar)\simeq 1.3\times10^{18}\,{\rm V}/{\rm m}$ and magnetic $B_{\rm cr}=E_{\rm cr}/c\simeq4.4\times10^9\,{\rm T}$ fields.
For electric $E$ and magnetic $B$ fields fulfilling $E\ll E_{\rm cr}$ and $B\ll B_{\rm cr}$, it can be presumed that nonlinear corrections to classical electrodynamics in vacuo governed by the Maxwell Lagrangian ${\cal L}_{\rm MW}=-\frac{1}{4}F_{\mu\nu}F^{\mu\nu}=\frac{1}{2}(E^2-B^2)$ are suppressed with powers of $E/E_{\rm cr}$ and $B/B_{\rm cr}$, respectively.
All macroscopic electromagnetic fields available in the laboratory, even those provided by state-of-the-art high-intensity laser systems reaching peak field amplitudes of the order of $E\simeq10^{14}\,{\rm V}/{\rm m}$ and $B\simeq10^6\,{\rm T}$, meet the criterion of $\{E/E_{\rm cr},B/B_{\rm cr}\}\ll1$.

Contributions of other particle sectors of the Standard Model are suppressed even more as their masses $m$ fulfill $m\gg m_e$ while their charges are of the same order as $e$, implying the associated critical fields to be even larger.
Note that at first sight the tiny masses of free quarks seem to contradict this statement. However, quarks directly couple to gluons and their physics is governed by quantum chromodynamics (QCD). Central properties of QCD are color confinement and infrared slavery, which implies strong coupling at low energies. Hence, it is to be expected that upon integrating out the gluons, the spectrum of QCD at low energies is characterized by colorless states made up of quarks and gluons. Their lightest representatives are the pions, which could in turn mediate effective interactions between prescribed electromagnetic fields and clearly fulfill $m\gg m_e$.

In this article we aim at exploring the perspectives of inducing a sizable QED vacuum birefringence signal at the Gamma Factory. Gamma Factory is a proposed research infrastructure at CERN \cite{Krasny:2015ffb,Budker:2020zer} included in the Physics Beyond Colliders programme \cite{Jaeckel:2020dxj}.
It is intended to deliver gamma photons with energies up to about $400\,{\rm MeV}$ at high photon fluxes emitted into a narrow cone. The gamma photons are envisioned to be produced via the resonant scattering of laser photons by highly relativistic, partially-stripped ions circulating in the accelerator.

Our article is structured as follows: after establishing the parameter regime for which photon propagation effects in the QED vacuum subjected to a macroscopic electromagnetic background field can be reliably analyzed on the basis of the leading contribution to the Heisenberg-Euler effective Lagrangian in Sec.~\ref{sec:quantumvacuum}, we focus on the study of vacuum birefringence in Sec.~\ref{sec:vacbiref}. Using the gamma photons provided by the Gamma Factory as probe, we analyze two different scenarios: vacuum birefringence driven by magnetic field in Sec.~\ref{sec:magnfield}, and by a high-intensity laser field in Sec.~\ref{sec:hilfield}.
Finally, we end with Conclusions in Sec.~\ref{sec:concls}.
Throughout this work, we use the Heaviside-Lorentz system and units with $c=\hbar=1$.

\section{Quantum vacuum effects}\label{sec:quantumvacuum}

Our next goal is to construct the effective theory describing the dynamics and interactions of macroscopic electromagnetic fields in the quantum vacuum. Here, we focus on the limit of  $\{E/E_{\rm cr},B/B_{\rm cr}\}\ll1$ relevant for laboratory experiments; cf. the recent reviews~\cite{DiPiazza:2011tq,Battesti:2012hf,King:2015tba,Karbstein:2019oej} and references therein.
Correspondingly, a perturbative expansion of the vacuum-fluctuation-mediated effective interactions of the applied electromagnetic fields in powers of $E$ and $B$, or -- resorting to a covariant and gauge invariant formulation -- in powers of the field strength tensor $F^{\mu\nu}$ and its dual  ${}^*\!F^{\mu\nu} =\frac{1}{2}\epsilon^{\mu\nu\rho\sigma}F_{\rho\sigma}$, can be truncated at low orders; $\epsilon^{\mu\nu\rho\sigma}$ is the Levi-Civita symbol.

As a consequence of the fact that QED is CP invariant, i.e., respects a charge conjugation parity symmetry, we also know that these effective interactions are even in both $F$ and ${}^*\!F$. Another building block for constructing such effective interactions are derivatives $\partial^\mu$. The latter are rendered dimensionless by the electron mass $m_e$, which suggests that derivative corrections may be neglected if the applied electromagnetic fields vary on scales much larger than the Compton wavelength of the electron $\lambdabar_{\rm C}=\hbar /(m_ec)\simeq3.8\times10^{-13}\,{\rm m}$.
While we will come back to this point later, let us for the moment just assume the possibility of a perturbative expansion in powers of the derivative $\partial$ and resort to a derivative expansion.
For convenience, we define dimensionless derivatives $\hat\partial=\partial/m_e$ and field strengths $\hat F =eF/m_e^2$ and ${}^*\!\hat{F}=e{}^*\!F/m_e^2$. These notations are only used briefly: they are intended to streamline the discussions in Secs.~\ref{sec:EFT} and \ref{sec:LPP}.

\subsection{Effective Lagrangian}\label{sec:EFT}

In turn, we can infer the following structure of the vacuum-fluctuation-mediated effective interactions of the applied electromagnetic fields ${\cal L}_{\rm int}$ supplementing ${\cal L}_{\rm MW}$ with quantum corrections,
\begin{align}
 {\cal L}_{\rm int}/m_e^4=\,& \bigl(a_2\hat\partial\hat\partial+a_4\hat\partial\hat\partial\hat\partial\hat\partial+\ldots\bigr)\hat F\hat F \nonumber\\
&\ +\bigl(b_0+b_2\hat\partial\hat\partial+b_4\hat\partial\hat\partial\hat\partial\hat\partial+\ldots\bigr)\hat F\hat F\hat F\hat F + \bigl(c_0+c_2\hat\partial\hat\partial+c_4\hat\partial\hat\partial\hat\partial\hat\partial+\ldots\bigr)\hat F\hat F {}^*\!\hat F{}^*\!\hat F \nonumber\\
 &\ +\bigl(d_0+d_2\hat\partial\hat\partial+d_4\hat\partial\hat\partial\hat\partial\hat\partial+\ldots\bigr)\hat F\hat F\hat F\hat F\hat F\hat F + \ldots\,, \label{eq:Lint}
\end{align}
where $a_i$, $b_i$, $c_i$ and $d_i$ denote numerical coefficients; the index $i$ counts the numbers of derivatives. Some clarifications are in order here. As the Lagrangian is a Lorentz scalar, the Minkowski indices of each single contribution in \Eqref{eq:Lint} are to be fully contracted. Accounting for the fact that $\hat F$ and ${}^*\!\hat{F}$ are second-rank tensors, this immediately implies that Eq.~\eqref{eq:Lint} is also even in the derivatives.
Moreover, a specific term in \Eqref{eq:Lint} represents the set of all non-redundant contractions of its constituents multiplied by individual numeric coefficients. For example, $b_2\hat\partial\hat\partial\hat F\hat F\hat F\hat F$ serves as representative for all possible independent contractions of 4 field strength tensors with two derivatives and their prefactors. Analogously, $c_2\hat\partial\hat\partial \hat F\hat F {}^*\!\hat F{}^*\!\hat F $ stands for all independent contractions of two $\hat F$, two ${}^*\!\hat{F}$ and two derivatives which were not already accounted for in the previous set.

We also emphasize that this formal expansion of course does not have to be convergent. Instead, the numerical coefficients may grow rapidly with increasing orders in the expansions. In fact, expansions in the number of derivatives and powers of the field are typically non-convergent but asymptotic \cite{Dunne:2004nc}.
Besides, we note that by construction a perturbative expansion in the field strength is of course insensitive to manifestly nonperturbative phenomena, such as the principle possibility of electron-positron pair-production via the Schwinger effect \cite{Sauter:1931zz,Heisenberg:1935qt,Schwinger:1951nm}. These effects can, however, be safely neglected in the parameter regime to be considered below; cf., e.g., Refs.~\cite{Tsai:1974fa,Dittrich:1985yb,Dittrich:2000zu,Dunne:2004nc,Karbstein:2013ufa} and references therein.

\subsection{Light propagation phenomena}\label{sec:LPP}

In a next step, we decompose the field $F$ as $F\to F+f$ into a pump $F$ and a probe light field $f$. We then limit ourselves to the contribution bilinear in the probe field, which we denote by ${\cal L}_{{\rm int}}^{ff}$. The latter allows for the study of the effect of quantum corrections on light propagation phenomena, characterized by both an incident and an outgoing light field $f$. Correspondingly, we have
\begin{align}
 {\cal L}_{{\rm int}}^{ff}/m_e^4=\,& \bigl(a_2\hat\partial\hat\partial+a_4\hat\partial\hat\partial\hat\partial\hat\partial+\ldots\bigr)\hat f\hat f \nonumber\\
&\ +\bigl(b_0+b_2\hat\partial\hat\partial+b_4\hat\partial\hat\partial\hat\partial\hat\partial+\ldots\bigr)\hat f\hat f\hat F\hat F + \ldots \nonumber\\
 &\ +\bigl(d_0+d_2\hat\partial\hat\partial+d_4\hat\partial\hat\partial\hat\partial\hat\partial+\ldots\bigr)\hat f\hat f\hat F\hat F\hat F\hat F + \ldots\,.
 \label{eq:Lintbilin}
\end{align}
Here, $b_2\hat\partial\hat\partial\hat f\hat f\hat F\hat F$ denotes all non-redundant contractions of two derivatives, two $f$ and two $F$ with individual coefficients, etc. To keep \Eqref{eq:Lintbilin} simple, we refrain from explicitly stating the contributions arising from the second term in the second line of \Eqref{eq:Lint}.

Being interested in the study of light propagation effects at weak field strengths, subsequently we assume the probe field $f$ (vector potential $a^\mu$) to be paraxial-like and approximately transverse: it features a well-defined propagation direction $\hat{\vec{\kappa}}$ along which it propagates essentially with the speed of light in vacuum. At the same time, its typical frequency of variation is given by its oscillation frequency $\omega$.
In turn, its dominant variation is scaling as $f(x)\sim \omega a(x)\sim{\rm e}^{{\rm i}kx}$, with $k^\mu\approx\omega(1,\hat{\vec{\kappa}})$ and $k^2=k_\mu k^\mu\approx0$. Approximate transversality means $k_\mu  f^{\mu\nu}=(kf)^\nu\approx0$ and $(k{}^*\!f)^\nu\approx0$.
On the other hand, the pump field $F$ does not necessarily correspond to a propagating light field, but could be a generic electromagnetic field. However, we assume its typical frequency scale of variation $\Omega$ to be constrained by $\Omega/m_e\ll1$.
The latter criterion immediately implies that all contributions in \Eqref{eq:Lintbilin} involving derivatives acting exclusively on $\hat F$ or ${}^*\!\hat F$ can be dropped from the outset. Due to the fact that $f$ is approximately light-like and transverse, also all derivatives contracted with itself $\hat\partial^2=\hat\partial_\mu\hat\partial^\mu$ which act on $\hat f$ or ${}^*\!\hat f$, as well as derivatives being both contracted with and acting on either $\hat f$ or ${}^*\!\hat f$ can be safely neglected.
Besides, derivatives contracted with itself which simultaneously act on a representant of $\hat F$, ${}^*\!\hat F$ and $\hat f$, ${}^*\!\hat f$, such as $\hat\partial_\mu \hat f \hat\partial^\mu \hat{F}$, result in factors of $\omega\Omega/m_e^2$.

The contributions in the second line of \Eqref{eq:Lintbilin} are quadratic in the field strength tensors $\hat F$ and ${}^*\!\hat{F}$ of the pump field. As these are antisymmetric in their Minkowski indices, two derivatives contracted to the same field strength tensor vanish identically.
In turn, at most two derivatives can be contracted with these tensors: namely one derivatives with each of the $\hat F$.
Analogously, at most $2n$ derivatives can be contracted with the pump field strength tensors constituting the contribution containing $2n$ factors of $\hat F$ and ${}^*\!\hat{F}$. Each of these derivatives $\hat\partial$ acting on $\hat f$ yields a factor proportional $\omega/m_e$, while a derivative acting on $\hat F$ or ${}^*\!\hat F$ generically comes with a factor of $\Omega/m_e$.

If the probe frequency fulfills the same criterion as the pump frequency, namely $\omega/m_e\ll1$, anyhow all derivative terms can be dropped from the outset.
On the other hand, the above discussion implies that for $\omega\gtrsim m_e$, while $\omega\Omega/m_e^2\ll1$, the dominant contribution to \Eqref{eq:Lintbilin} at $(2n)$th order in the pump field scales at most as $(\omega/m_e)^{2n}$.
In fact, we have established that the contribution which may scale as $(\omega/m_e)^{2n}$ is of the structure 
\begin{equation}
 \hat f\hat f\underbrace{(\hat k\hat F)\ldots(\hat k\hat F)}_{n\ \text{times}} \,, \label{eq:omega^n}
\end{equation}
where $\hat k=k/m_e$ and $(\hat k\hat F)^\nu=\hat k_\mu F^{\mu\nu}$.
An analysis of the possible contractions of the Minkowski indices in \Eqref{eq:omega^n} unveils that this contribution vanishes approximately because of $k^2\approx 0$, $(kf)^\nu\approx0$ and $(k{}^*\!f)^\nu\approx0$.
Hence, in addition accounting for the scaling with the field strength of the pump field and using that $\hat{f}\sim(\omega/m_e)\hat{a}$, we find that for $\omega\Omega/m_e^2\ll1$ the term in \Eqref{eq:Lintbilin} containing $2n$ factors of $\hat F$ and ${}^*\!\hat{F}$ scales at most as $(e{\cal E}\omega/m_e^3)^{2n} \hat{a}\hat{a}$, where we introduced ${\cal E}=\max\{E,B\}$.
This suggests that all contributions beyond $n=1$ in \Eqref{eq:Lintbilin} can be safely neglected as long as the condition $(e{\cal E}\omega/m_e^2)^2\ll1$ holds.

Correspondingly, the criteria for the leading term containing no derivatives in \Eqref{eq:Lint},
\begin{align}
 {\cal L}_{\rm int}/m_e^4\simeq b_0\hat F\hat F\hat F\hat F +c_0\hat F\hat F {}^*\!\hat F{}^*\!\hat F \,,
 \label{eq:LintLO}
\end{align}
to allow for the reliable theoretical study of the impact of quantum vacuum nonlinearities on light propagation phenomena in weak, slowly varying background fields (field strength $F$, frequencies $\Omega$) characterized by the conditions
\begin{equation}
 \Bigl(\frac{eF}{m_e^2}\Bigr)^2\ll1\quad\text{and}\quad\Bigl(\frac{\Omega}{m_e}\Bigr)^2\ll1\,, \label{eq:criterion0}
\end{equation}
are approximately on-shell and transverse probe light fields (oscillation frequency $\omega$) fulfilling
\begin{equation}
 \Bigl(\frac{eF\omega}{m_e^3}\Bigr)^2\ll1\quad\text{and}\quad\frac{\omega\Omega}{m_e^2}\ll1\,. \label{eq:criterion}
\end{equation}

We in particular emphasize that \Eqref{eq:criterion} does not limit the probe frequency to be much smaller than the electron mass, as might be naively expected from the fact that the relevant reference scale rendering the derivative dimensionless is $m_e$; see the corresponding discussion before \Eqref{eq:Lint}. Instead, it implies that as long as $eF/m_e^2\ll1$, the ratio $\omega/m_e$ can obviously be much larger than unity, the relevant criterion being $\omega/m_e\ll{\rm min}\{(eF/m_e^2)^{-1},(\Omega/m_e)^{-1}\}$.

Analogous considerations are possible for other signatures of quantum vacuum nonlinearity in background fields such as, e.g., probe photon splitting and merging processes encoded in ${\cal L}_{\rm int}^{fff}$, etc.

For completeness, we note that the conditions \eqref{eq:criterion0} and \eqref{eq:criterion} are fully compatible with those which can be inferred from the exact results for the one-loop photon polarization tensors evaluated in constant and plane-wave background fields \cite{Tsai:1975iz,Heinzl:2006pn,Karbstein:2013ufa,Gies:2014jia}. However, the current derivation is more general: it neither requires the background field to be constant, nor of plane-wave type, but applies to light propagation phenomena in generic slowly varying electromagnetic fields.

The explicit expression of \Eqref{eq:LintLO} accounting for the correct numeric coefficients at one-loop order is \cite{Euler:1935zz,Heisenberg:1935qt}
\begin{align}
 {\cal L}_{\rm int}/m_e^4\simeq\frac{1}{360\pi^2}\Bigl(\frac{e}{m_e^2}\Bigr)^4
 \Bigl(4{\cal F}^2+7{\cal G}^2\Bigr)\,.
 \label{eq:LintLOexpl}
\end{align}
with
\begin{equation}
{\cal F}=\frac{1}{4}F_{\mu\nu}F^{\mu\nu}\quad\text{and}\quad{\cal G}=\frac{1}{4}F_{\mu\nu}{}^*\!F^{\mu\nu}\,.
\end{equation}
This expression amounts to the leading contribution of the renowned (one-loop) Heisenberg-Euler effective Lagrangian~\cite{Heisenberg:1935qt}. It gives rise to the effective coupling of four electromagnetic fields mediated by quantum vacuum fluctuations of electrons and positrons.
Higher loop corrections are parametrically suppressed with powers of $\alpha=e^2/(4\pi)\simeq1/137$; see Ref.~\cite{Ritus:1975cf} for the coefficient in \Eqref{eq:LintLO} at two loops. Hence, given that the above conditions~\eqref{eq:criterion0} and \eqref{eq:criterion} are met, \Eqref{eq:LintLOexpl} should allow for the accurate theoretical study of light propagation phenomena in the presence of experimentally realistic macroscopic electromagnetic fields. However, particularly due to the non-convergent nature of the employed expansions (cf. Sec.~\ref{sec:EFT}), sizable deviations are to be expected when any of the parameters in Eqs.~\eqref{eq:criterion0} and \eqref{eq:criterion} becomes of the order of $\lesssim1$.

\section{Vacuum Birefringence}\label{sec:vacbiref}

\begin{figure}
 \includegraphics[width=0.25\linewidth]{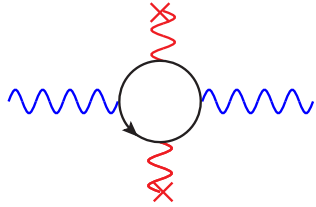}
 \caption{Leading order Feynman diagram giving rise to vacuum birefringence. A prescribed electromagnetic field (red wiggly lines ending at crosses) can induce a birefringence phenomenon for traversing probe photons (blue wiggly lines). The effective coupling of the probe and background fields is mediated by an electron-positron vacuum fluctuation (black solid line).}
 \label{fig:feynman}
\end{figure}

Subsequently, we focus on the experimental signature of vacuum birefringence \cite{Toll:1952rq,Klein:1964,Baier:1967zzc,Baier:1967zza,BialynickaBirula:1970vy,Brezin:1971nd,Adler:1971wn} arising from the effective non-linear interaction of the electromagnetic fields in \Eqref{eq:LintLOexpl}.  See Fig.~\ref{fig:feynman} for the corresponding Feynman diagram. As the signature of vacuum birefringence is expected to scale with a positive power of the energy of the probe photons, the use of gamma photons looks particularly promising.
This effect is already actively searched for in experiments employing continuous wave lasers in combination with high-finesse cavities as probe and macroscopic, quasi-constant and quasi-static magnetic fields of a few Tesla to induce the effect \cite{Battesti:2018bgc,Ejlli:2020yhk,Agil:2021fiq,Fan:2017fnd} resorting to the scheme devised by Ref.~\cite{Iacopini:1979ci}.
See Refs.~\cite{Aleksandrov:1985,Kotkin:1996nf,Luiten:2004pz,Luiten:2004py,Heinzl:2006xc,DiPiazza:2006pr, Karbstein:2018omb,Dinu:2013gaa,Dinu:2014tsa,Karbstein:2015xra,Ilderton:2016khs,Karbstein:2016lby,Schlenvoigt:2016jrd,Nakamiya:2015pde,King:2016jnl,Bragin:2017yau,Ataman:2018ucl,Kadlecova:2019dxv,Tzenov:2019ovq,Robertson:2020nnc} for proposals aiming at probing the effect with optical, x-ray, synchrotron or gamma radiation and high-intensity lasers as pump. For proposals to measure vacuum birefringence induced by a magnetic field with high-energy photons, cf. Ref.~\cite{Cantatore:1991sq,Wistisen:2013waa}.

As detailed below, it turns out that the parameter regimes readily accessible with the Gamma Factory can be reliably analyzed on the basis of the leading contribution to the Heisenberg-Euler effective Lagrangian in Eq.~\eqref{eq:LintLOexpl}.
We reiterate that the big advantage of the parameter regime constrained by Eqs.~\eqref{eq:criterion0} and \eqref{eq:criterion} is the possibility of studying the phenomenon in manifestly inhomogeneous, experimentally realistic pump and probe field configurations, such as provided by focused laser pulses.
Beyond this regime a controlled analytical study of light propagation phenomena is only possible in the special cases of homogeneous constant electromagnetic fields and plane-wave backgrounds for which the one-loop photon polarization tensor is known explicitly \cite{Batalin:1971au,Becker:1974en,Baier:1974hn,Baier:1975ff}.

\subsection{Magnetic field}\label{sec:magnfield}

[To be specific, here we study the effect in an unidirectional, quasi-constant and static magnetic field. 
For simplicity, we assume the frequency-$\omega$ probe beam to be well-described as a linearly polarized transverse plane wave propagating along the positive $\rm x$ axis. Correspondingly, the associated magnetic and electric fields are $\vec{b}=E\,\vec{e}_\beta$ and $\vec{e}=E\,\vec{e}_{\beta+\frac{\pi}{2}}$, with amplitude profile $E=E(x)=E_0\cos\bigl(\omega(t-{\rm x})\bigr)$ and $\vec{e}_\beta=(0,\sin\beta,\cos\beta)$; $E_0$ is the peak field amplitude. The choice of the angle $\beta$ fixes the polarization direction of the probe.
At the same time, the background magnetic field $\vec{\cal B}={\cal B}\vec{e}_{\rm z}$ is oriented along the $\rm z$ axis, 
and extends over a finite length in the propagation direction of the probe; transversally it is assumed to be constant. As long as the applied magnetic field is much wider than the diameter of the probe beam and essentially does not vary on this scale, the effectively $1+1$ dimensional calculation performed here can also be employed for a probe beam of finite spatial extent.

To study the effect it is convenient to resort to the vacuum emission picture~\cite{Galtsov:1971xm,Karbstein:2014fva}, providing an intuitive and straightforward framework for the study of photonic signatures of quantum vacuum nonlinearity in inhomogeneous electromagnetic fields.
In this approach \cite{Fritzsche:2019} the superposition of the driving electromagnetic fields gives rise to a signal-photon current, supplementing the outgoing fields with a signal-photon component constituting the signature of quantum vacuum nonlinearity.
In momentum space the latter can be expressed as
\begin{equation}
 j_\nu(k)={\rm i}\int{\rm d}^4x\,{\rm e}^{-{\rm i}kx} \Bigl(k^\mu F_{\mu\nu}\frac{\partial{\cal L}_{\rm int}}{\partial{\cal F}}+ k^\mu{}^\star F_{\mu\nu}\frac{\partial{\cal L}_{\rm int}}{\partial{\cal G}}\Bigr)\,, \label{eq:jk}
\end{equation}

Light propagation phenomena such as vacuum birefringence are encoded in contributions linear in the incident probe field; vacuum-fluctuation-mediated corrections to probe photon propagation manifest themselves in signal photons supplementing the outgoing probe field with quantum corrections.
Upon limitation to the leading effective interaction in \Eqref{eq:LintLOexpl}, the current~\eqref{eq:jk} is cubic in the driving field. This immediately implies the vacuum birefringence phenomenon studied here to be linear in the plane wave probe and quadratic in the background magnetic field.
Besides, the signal photon current in position space $j_\nu(x)$ is manifestly real-valued.
Being interested in the signal to be detected far outside the interaction region at vanishing magnetic field $B$, the electric field of the induced signal can then be expressed as \cite{Karbstein:2019oej}
\begin{equation}
 \vec{e}_{\rm s}(x)={\rm Re}\int\frac{{\rm d}^3k}{(2\pi)^3}\,{\rm e}^{{\rm i}kx}\,\underbrace{\bigl(\vec{j}(k)-\hat{\vec{k}}j^0(k)\bigr)}_{=:\vec{e}_{\rm s}(k)}\Big|_{k^0=|\vec{k}|}\,, \label{eq:ex}
\end{equation}
in terms of a Fourier transform of components of the current~\eqref{eq:jk} contracted with the signal photon momentum $k^\mu$ fulfilling the on-shell condition $k^2=0$; the associated unit momentum vector is $\hat{\vec{k}}=\vec{k}/{\rm k}$, where ${\rm k}=|\vec{k}|$.
Upon plugging Eqs.~\eqref{eq:LintLOexpl} and \eqref{eq:jk} into \Eqref{eq:ex}, the momentum space representation of the signal electric field can be cast in the following form,
\begin{align}
 \vec{e}_{\rm s}(k) \big|_{k^0=|\vec{k}|} \simeq\frac{1}{\rm i} \frac{{\rm k}m_e^4}{180\pi^2}\Bigl(\frac{e}{m_e^2}\Bigr)^4\int{\rm d}^4x\,{\rm e}^{-{\rm ik}(\hat{\vec{k}}\cdot\vec{x}-t)}\Big(&4\bigl[\vec{E}-\hat{\vec{k}}(\hat{\vec{k}}\cdot\vec{E})+\hat{\vec{k}}\times\vec{B}\bigr]{\cal F} \nonumber\\
 +&7\bigl[ \vec{B}-\hat{\vec{k}}(\hat{\vec{k}}\cdot\vec{B}) -\hat{\vec{k}}\times\vec{E}\bigr]{\cal G}\Bigr)\,. \label{eq:ek}
\end{align}
In the present case, we obviously have $\vec{B}={\cal B}\vec{e}_{\rm z}+E\,\vec{e}_\beta$ and $\vec{E}=E\,\vec{e}_{\beta+\frac{\pi}{2}}$, such that ${\cal F} = \frac{1}{2}{\cal B}^2+{\cal B}E \cos\beta$ and ${\cal G} = {\cal B}E\sin\beta$.
Keeping terms linear in $E_0$ only, \Eqref{eq:ek} can be rewritten as
\begin{align}
 \vec{e}_{\rm s}(k) \big|_{k^0=|\vec{k}|} \simeq\frac{1}{\rm i} \frac{\alpha}{\pi}\frac{1}{90}E_0&\big[2(\omega-3 k_{\rm x})\vec{e}_{\rm y}\cos\beta +(5\omega+2k_{\rm x})\vec{e}_{\rm z} \sin\beta\bigr] \nonumber\\ \times&(2\pi)^3\delta(k_{\rm y}) \delta(k_{\rm z})\delta(|k_{\rm x}|-\omega)\int{\rm dx}\, {\rm e}^{-{\rm i}(k_{\rm x}-\omega){\rm x}} \,\biggl(\frac{e{\cal B}({\rm x})}{m_e^2}\biggr)^2\,, 
\end{align}
where all trivial Fourier integrations have been performed.

Taking into account the physical boundary condition that the far-field signal evaluated here emerges from the localized space-time region where the driving fields overlap, the electric field characterizing the signal in position space~\eqref{eq:ex} decomposes into two distinct contributions: a signal co-propagating ``$+$'' with the incident probe beam,
\begin{align}
 \vec{e}^{\,\,+}_{\rm s}(x)\simeq -E_0\sin\bigl(\omega({\rm x}-t)\bigr)\bigl(4\vec{e}_{\rm y}\cos\beta -7\vec{e}_{\rm z} \sin\beta\bigr)\, \frac{\alpha}{\pi} \frac{1}{90}\,\omega\int{\rm dx}'\,\biggl(\frac{e{\cal B}({\rm x}')}{m_e^2}\biggr)^2\,, \label{eq:ex+}
\end{align}
and another one induced in the opposite ``$-$'' direction,
\begin{align}
 \vec{e}^{\,\,-}_{\rm s}(x)\simeq -E_0\sin\bigl(\omega({\rm x}+t)\bigr)&\bigl(8\vec{e}_{\rm y}\cos\beta +3\vec{e}_{\rm z} \sin\beta\bigr)\, \frac{\alpha}{\pi} \frac{1}{90}\,\omega\int{\rm dx}'\, {\rm e}^{{\rm i}2\omega{\rm x}'} \biggl(\frac{e{\cal B}({\rm x}')}{m_e^2}\biggr)^2\,. \label{eq:ex-}
\end{align}
The former contribution~\eqref{eq:ex+} is proportional to the integral of the squared magnetic field profile along the optical path of the probe, $\int{\rm dx}'\,{\cal B}^2({\rm x}')$.
On the other hand, \Eqref{eq:ex-} scales as the Fourier transform of the magnetic field squared to be evaluated at double the probe frequency, $\int{\rm dx}'\,{\rm e}^{{\rm i}2\omega{\rm x}'}{\cal B}^2({\rm x}')$. This contribution can be traced back to a quantum reflection process \cite{Gies:2013yxa}.
If the typical frequency scales of variation of the background field $\Omega$ fulfill the criterion $\Omega\ll\omega$, as will be the case in the specific scenario considered below, we obviously have $\int{\rm dx}'\,{\rm e}^{{\rm i}2\omega{\rm x}'}{\cal B}^2({\rm x}')\to0$.
Therefore, only \Eqref{eq:ex+} needs to be accounted for in the following discussion.

Upon superposition with the electric field of the probe beam traversing the interaction region with the pump field essentially unmodified, we obtain an expression for the outgoing electric field in forward direction,
\begin{align}
 \vec{E}^+(x)&\simeq E_0 \cos\bigl(\omega(t-{\rm x})\bigr) \bigl(\vec{e}_{\rm y}\cos\beta-\vec{e}_{\rm z}\sin\beta\bigr)\nonumber\\
 &\quad-E_0\sin\bigl(\omega({\rm x}-t)\bigr)\bigl(4\vec{e}_{\rm y}\cos\beta -7\vec{e}_{\rm z} \sin\beta\bigr)\, \underbrace{\frac{\alpha}{\pi}\frac{1}{90}\,\omega\int{\rm dx}'\,\biggl(\frac{e{\cal B}({\rm x}')}{m_e^2}\biggr)^2}_{=:\Delta}\,. \label{eq:Ex+}
\end{align}
For generic choices of the polarization vector of the incident probe light $\beta$, this field is elliptically polarized.
For the maximum and minimum values for the modulus squared of \Eqref{eq:Ex+} be find
\begin{align}
	| \vec{E}^+(x) |^2\big|_{\rm max}&\simeq E_0^2\bigl(1+{\cal O}(\Delta^2)\bigr)\,, \nonumber\\
	| \vec{E}^+(x) |^2\big|_{\rm min}&\simeq E_0^2\Delta^2\Bigl(\frac{9}{4}\sin^2(2\beta)+{\cal O}(\Delta^2)\Bigr)\,. \label{eq:ellaxes}
\end{align}
The maximum and minimum values are separated by a phase difference of $\pi/2$.
Here, we only kept the leading terms in a perturbative expansion in the dimensionless quantity $\Delta$ defined in \Eqref{eq:Ex+}. Note, that the consistent determination of the contributions scaling as $\Delta^{2(n-1)}$ in \Eqref{eq:ellaxes} would require us to account for effective interactions up to $2n$th order in $F$ to ${\cal L}_{\rm int}$; with \Eqref{eq:LintLOexpl} we are explicitly limited to $n=2$.

In turn, with the accuracy of the terms given explicitly in \Eqref{eq:ellaxes} the associated ellipticity angle $\chi$ is given by
\begin{equation}
	\chi=\arctan\biggl(\frac{| \vec{E}^+(x) |\big|_{\rm min}}{| \vec{E}^+(x) |\big|_{\rm max}}\biggr)\simeq\frac{3}{2}\Delta|\sin(2\beta)|\,. \label{eq:chi}
\end{equation}
Obviously, the ellipticity vanishes for $\beta=0\ ({\rm mod}\ \pi/2)$, and is maximized for $\beta =\pi/4\ ({\rm mod}\ \pi/2)$.
The phase difference $\Phi$ accumulated by the probe beam after traversing the pump magnetic field is \cite{Jekrard:1954}
\begin{equation}
	\Phi=2\chi\simeq  \frac{\alpha}{\pi}\frac{1}{30} \sin(2\beta)\,\omega\int{\rm dx}'\,\biggl(\frac{e{\cal B}({\rm x}')}{m_e^2}\biggr)^2\,. \label{eq:Phi}
\end{equation}
For a magnetic field fulfilling $\int{\rm dx}'\,{\cal B}^2({\rm x}')=l{\cal B}_0^2$, which resembles a constant magnetic field of strength ${\cal B}_0$ extending over a length $l$, we recover the well-known constant field result $\Phi\simeq\omega l\alpha/(30\pi)(e{\cal B}_0/m_e^2)^2\sin(2\beta)$. This result implies that the phase difference increases linearly with both the photon energy of the probe and the length of the applied magnetic field, but quadratically with the field strength of the latter; cf. Fig.~\ref{fig:feynman}.

At the same time, the birefringence property of the polarized quantum vacuum results in signal photons scattered into a mode polarized perpendicularly to the incident probe light of electric field $\vec{E}(x)=E(x)\vec{e}_{\beta+\frac{\pi}{2}}$.
As the latter is polarized along $\vec{e}_\parallel:=\vec{e}_{\beta+\frac{\pi}{2}}$, the polarization vector of the former is $\vec{e}_\perp:=\vec{e}_{\beta}$.
A comparison of the results in \Eqref{eq:ellaxes} with the electric field of the incident probe implies that $\vec{E}^+(x)\big|_{\rm max}\sim \vec{e}_\parallel$ and $\vec{E}^+(x)\big|_{\rm min}\sim \vec{e}_\perp$.
Correspondingly, we have $\vec{E}^+(x)\cdot\vec{e}_\parallel\simeq E(x)$ and $\vec{E}^+(x)\cdot\vec{e}_\perp=: E_\perp(x)$, such that
\begin{equation}
	\frac{N_\perp}{N}\simeq\frac{\int{\rm d}t\int{\rm d}A\,\langle\vec{E}_\perp^2(x)\rangle}{\int{\rm d}t\int{\rm d}A\,\langle\vec{E}^2(x)\rangle}\,, \label{eq:NperpbyN}
\end{equation}
where $N_\perp$ denotes the number of polarization-flipped signal photons, and $N$ the number of photons constituting the probe. The integrations over time and transverse area are assumed to be carried out at a fixed longitudinal coordinate in the far field.
For the specific case of a plane wave probe traversing an effectively one-dimensional field inhomogeneity without transverse structure as considered here, \Eqref{eq:NperpbyN} immediately implies that $N_\perp/N\simeq | \vec{E}^+(x) |^2\big|_{\rm min} /| \vec{E}^+(x) |^2\big|_{\rm max}$. Correspondingly, we have $N_\perp/N\simeq(\Phi/2)^2$; cf. also Ref.~\cite{Karbstein:2015qwa} which directly determined the number of perpendicularly polarized photons for the scenario considered here without resorting to a determination of the phase difference.

\begin{figure}
 \includegraphics[width=0.6\linewidth]{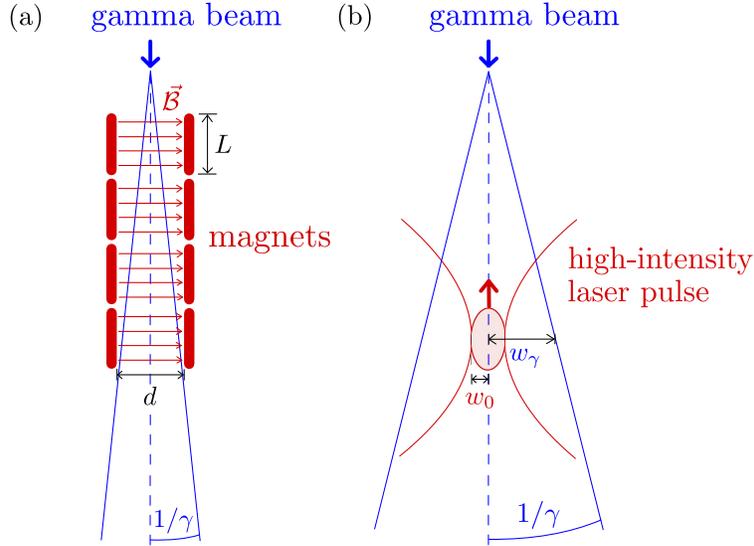}
 \caption{Graphical depiction of the two scenarios envisioned for the study of QED vacuum birefringence at the Gamma Factory discussed in the present article. In case (a) the birefringence phenomenon is induced by the quasi-constant static magnetic field provided by LHC dipole magnets. In case (b) the effect is driven by a counter-propagating focused high-intensity laser pulse.}
 \label{fig:variants}
\end{figure}

The Gamma Factory \cite{Krasny:2015ffb,Budker:2020zer} will enable such a vacuum birefringence experiment with a probe photon energy as high as $\omega\simeq400\,{\rm MeV}$. The driving magnetic field could be provided by a sequence of LHC dipole magnets, providing a magnetic field of strength ${\cal B}_0\simeq8.3\,{\rm T}$ over a length of $l\simeq14.3\,{\rm m}$ each \cite{Savary:2008zz}; the effective diameter $d$ of the bore for traversing light is about $d\simeq45\,{\rm mm}$.
In this case, we obviously have $\{(\Omega/m_e)^2,\omega\Omega/m_e^2\}\ll1$ as well as $(e{\cal B}_0/m_e^2)^2\simeq 3.54\times10^{-18}$ and $(e{\cal B}_0\omega/m_e^3)^2\simeq2.17\times10^{-12}$
fully compatible with the conditions~\eqref{eq:criterion0} and \eqref{eq:criterion}; cf. also Refs.~\cite{Cantatore:1991sq,Wistisen:2013waa}.
To achieve gamma photon energies up to $\omega\simeq400\,{\rm MeV}$, the Lorentz factor $\gamma$ which effectively governs the generation of the high-energy gamma beam in the Gamma Factory needs to be as large as $\gamma\approx3000$. As the opening angle of the gamma beam is given by $\approx1/\gamma$, the bore diameter of the magnet immediately implies a maximum length $l_{\rm max}\approx (d/2)\gamma\simeq67.5\,{\rm m}$ of the magnetic field provided by LHC magnets through which the full gamma beam could travel.
In turn, we could envision the use of up to $4$ LHC dipole magnets resulting in $\omega l\simeq1.16\times10^{17}$; see Fig.~\ref{fig:variants}(a) for an illustration. As $1/\gamma\ll1$, the effectively $1+1$ dimensional analysis of the vacuum birefringence effect performed here, which neglects any transverse variations of the magnetic field, is still perfectly justified. Finally, in order to maximize the signal we choose $\beta=\pi/4$.

Equation~\eqref{eq:Phi} predicts that after traversing four LHC dipole magnets, the gamma beam has picked up a phase difference of \cite{Erber:1961}
\begin{equation}
	\Phi\simeq 4L\omega\sin(2\beta)\,\frac{\alpha}{\pi}\frac{1}{30}\Bigl(\frac{e{\cal B}_0}{m_e^2}\Bigr)^2\simeq3.18\times10^{-5}\,. \label{eq:PhiBconst}
\end{equation}
Note that this value is about an order of magnitude larger than the one predicted to be accessible in the head-on collision of state-of-the-art petawatt-class high-intensity-laser and free-electron-laser (FEL) pulses of $\omega\simeq{\cal O}(10)\,{\rm keV}$ \cite{Heinzl:2006xc,Karbstein:2018omb}.
At the same time, for the present parameters the fraction of incident gamma photons scattered into a perpendicularly polarized mode is given by $N_\perp/N\simeq2.53\times10^{-10}$.
For x-rays of $\omega\simeq{\cal O}(10)\,{\rm keV}$ the possibility of measuring such tiny ellipticities has been demonstrated experimentally \cite{Marx:2013xwa,Bernhardt:2020vxa,Schmitt:2020ttm}.
However, we emphasize that these techniques developed for x-ray polarimetry can certainly not be used at $400\,{\rm MeV}$. In the latter parameter regime rather pair-production polarimetry may be an option \cite{Eingorn:2018,Chattopadhyay:2021mbb}, but the possibility to measure tiny ellipticities such as the one in Eq.~\eqref{eq:PhiBconst} with this technique remains to be shown.
For proposals of prospective experimental schemes to measure high-energy vacuum birefringence utilizing pair production in matter to determine the polarization state of the outgoing probe photons, cf. Refs. \cite{Cantatore:1991sq, Wistisen:2013waa,Nakamiya:2015pde,Bragin:2017yau}.

The advantage of using a static magnetic field to drive the vacuum birefringence phenomenon is that essentially all gamma photons will traverse the magnetic field. In turn, the associated number of signal photons scattered in a perpendicularly polarized mode $N_\perp$ scales directly with the total number of gamma photons available for probing and thus is -- among other factors -- ultimately limited by the repetition rate of the gamma pulses.

\subsection{High-intensity laser field}\label{sec:hilfield}

Alternatively, the birefringence signal could be driven by a high-intensity laser field; see Fig.~\ref{fig:variants}(b) for an illustration.
In such a scenario, the birefringence signal is predominantly induced in the interaction region where the gamma probe collides with the focused high-intensity laser pulse reaching its peak field strength. Outside the focus, the field strength of the high-intensity pump rapidly drops. As the birefringence phenomenon scales as $\sim(e{\cal E}(x)/m_e^2)^2$ with the pump field strength at a given space-time coordinate $x$, the interaction of the probe with the substantially lower field outside the beam focus only gives rise to subleading corrections.

State-of-the-art high-intensity lasers of the petawatt-class typically deliver pulses of energy $W={\cal O}(10)\,{\rm J}$ and duration $\tau={\cal O}(10)\,{\rm fs}$ at a wavelength of $\lambda={\cal O}(1)\,\upmu{\rm m}$ and a repetition rate of ${\cal O}(1)\,{\rm Hz}$. These pulses can be focused to a waist radius of $w_0\gtrsim\lambda$.
In turn, the typical frequency scale of variation of the pump field is given by $\Omega=2\pi/\lambda$.
For our explicit example, we choose the parameters characterizing a readily available commercial 300 TW Titanium Sapphire laser system, such as the one installed at the Helmholtz International Beamline for Extreme Fields (HiBEF) at the European XFEL \cite{Hibef}:
$W=10\,{\rm J}$, $\tau=30\,{\rm fs}$, $\lambda=800\,\upmu{\rm m}$ and a repetition rate of $1\,{\rm Hz}$ focused to $w_0=1\,\upmu{\rm m}$.
Assuming the high-intensity laser field to be well-described as pulsed paraxial fundamental Gaussian beam, the electric peak field strength ${\cal E}_0$ in its focus can be expressed in terms of the pulse energy, pulse duration and waist radius as \cite{Karbstein:2017jgh}
\begin{equation}
 {\cal E}_0^2\simeq 8 \sqrt{\frac{2}{\pi}}\,\frac{W}{\pi w_0^2\tau}\,. \label{eq:peakfield}
\end{equation}
The associated Rayleigh range is ${\rm z}_{\rm R}=\pi w_0^2/\lambda$.

For the explicit parameters characterizing the high-intensity pump given above, the dimensionless quantities entering the conditions~\eqref{eq:criterion0} and \eqref{eq:criterion} are $(\Omega/m_e)^2\simeq9.2\times10^{-12}$, $(e{\cal E}_0/m_e^2)^2\simeq 1.46\times10^{-7}$, $\omega\Omega/m_e^2\simeq2.4\times10^{-3}$ and $(e{\cal E}_0\omega/m_e^3)^2\simeq0.09$. These values suggest that even in this parameter regime the size of the attainable vacuum birefringence signal can still be reliably estimated from the leading contribution to the Heisenberg-Euler effective Lagrangian in \Eqref{eq:LintLOexpl}.

Hence, for this particular scenario, the ratio of $| \vec{E}^+(x) |\big|_{\rm min} /| \vec{E}^+(x) |\big|_{\rm max}$ determining the ellipticity angle $\chi$ could be evaluated along the same lines as in the explicit calculation performed in Sec.~\ref{sec:magnfield} above.
This would result in a coordinate-dependent ellipticity angle \cite{DiPiazza:2006pr}. The ellipticity angle associated with the beam could, e.g., be defined in terms of the average of this ratio over the beam cross-section at a given longitudinal position in the far field.

On the other hand, the number of polarization-flipped signal photons $N_\perp$ populating the originally empty, perpendicularly polarized mode can be readily estimated with the analytic formula derived for the head-on collision of pump and probe laser pulses in Ref.~\cite{Karbstein:2018omb}; cf. in particular Eq.~(4) herein. In the present work, we will make use of this result for the ratio of $N_\perp/N$ and infer an estimate for the phase difference $\Phi$ picked up by the gamma beam after having been collided with the strong focused high-intensity laser pulse therefrom.

For vanishing spatio-temporal offsets, and an relative angle of $\beta$ between the polarization vectors of the pump and probe laser beams \cite{Karbstein:2019bhp}, this result can be represented as
\begin{equation}
 \frac{N_\perp}{N}\simeq\frac{4\alpha^4}{25(3\pi)^{3/2}}\Bigl(\frac{W}{m_e}\frac{\omega}{m_e}\sin(2\beta)\Bigr)^2\Bigl(\frac{\lambdabar_{\rm C}}{w_0}\Bigr)^4\, \frac{1}{1+2(\frac{w_\gamma}{w_0})^2}\,F\Bigl(\tfrac{\frac{4{\rm z}_{\rm R}}{T}}{\sqrt{1+\frac{1}{2}(\frac{\tau}{T})^2}},\tfrac{T}{\tau}\Bigr)\,, \label{eq:Nperp}
\end{equation}
with
\begin{equation}
	F(\chi,\rho)=\sqrt{\frac{1+2\rho^2}{3}}\,\chi^2\int_{-\infty}^\infty{\rm d}\kappa\,{\rm e}^{-\kappa^2(1+2\rho^2)}\biggl(\sum_{\ell=\pm1}{\rm e}^{(\ell\rho\kappa+\chi)^2}{\rm erfc}(\ell\rho\kappa+\chi)\biggr)^2\,.
\end{equation}
Here, ${\rm erfc}(.)$ is the complementary error function, $w_0$ and $w_\gamma$ are the radii of the pump and probe beams at the collision point, and $T$ is the pulse duration of the probe.
Given that the probe pulse duration fulfills $T\gg\{\tau,{\rm z}_{\rm R}\}$ and the beam radii at the collision point meet the criterion $w_\gamma\gg w_0$, \Eqref{eq:Nperp} simplifies significantly.
In this limit we have \cite{Karbstein:2018omb}
\begin{equation}
 \frac{1}{1+2(\frac{w_\gamma}{w_0})^2}\,F\Bigl(\tfrac{\frac{4{\rm z}_{\rm R}}{T}}{\sqrt{1+\frac{1}{2}(\frac{\tau}{T})^2}},\tfrac{T}{\tau}\Bigr)\simeq \frac{1}{2}\Bigl(\frac{w_0}{w_\gamma}\Bigr)^2\frac{\tau}{T}\,\sqrt{\frac{2\pi}{3}}\,\Bigl(\frac{8{\rm z}_{\rm R}}{\tau}\Bigr)^2\,{\rm e}^{(\frac{8{\rm z}_{\rm R}}{\tau})^2}\,{\rm erfc}\bigl(\tfrac{8{\rm z}_{\rm R}}{\tau}\bigr)\,. \label{eq:frac*F}
\end{equation}
The above criterion on the pulse duration should typically be fulfilled at the Gamma Factory \cite{Krasny:2015ffb,Budker:2020zer}, which can be expected to provide gamma pulses of duration $T\gtrsim160\,{\rm fs}$. At the same time, the collision point of the gamma beam with the high-intensity laser beam should be sufficiently separated from the source of the gamma photons such that generically $w_\gamma\gg w_0$.

We emphasize that this parameter regime even seems to be beneficial for high-intensity laser driven vacuum birefringence experiments: given that the conditions $T\gg\tau$ and $w_\gamma\gg w_0$ are met, the experiment is essentially insensitive to the shot-to-shot fluctuations inherent to high-intensity laser systems resulting in spatio-temporal offsets of ${\cal O}(w_0)$ of the position of the high-intensity laser focus. Variations of this order just change the location of the high-intensity laser focus within the forward cone of the gamma probe and thus do not impact the signal; cf. also Fig.~\ref{fig:variants}(b).

As already noted above, the opening angle of the gamma beam is given by $\theta\approx1/\gamma\ll1$, and thus can be safely neglected in the explicit evaluation of the signal in the interaction region. It only needs to be accounted for in the far-field where the signal is to be detected. However, even for a formally vanishing divergence of the incident probe the signal photons generically feature a finite far-field divergence because of the finite transverse extent of the interaction area of the colliding beams. If the beam radius $w_\gamma$ of the gamma beam in the interaction area with the high-intensity laser pulse fulfills $w_\gamma\gg w_0$, the latter scales as $\sim1/(\omega w_0)\ll1/\gamma$; see Eq.~(24) of Ref.~\cite{Karbstein:2018omb}. Correspondingly, in the present scenario the far-field angular divergence of the signal should be fully determined by geometric properties: the intercept theorem then immediately implies that the far-field radial divergence of the signal is $\theta_{\rm s}\approx w_0/(\gamma w_\gamma)$.
In line with that, in the far field the cross section $A_s$ of the signal relates to the cross section $A$ of the gamma beam as $A_s/A\approx(w_0/w_\gamma)^2$.
At the same time, generically the shorter one of the pulse durations $\tau$ and $T$ determines the pulse duration $\tau_{\rm s}$ of the signal. For $T\gtrsim\tau$, which is the regime of relevance here, we hence have $\tau_{\rm s}\approx\tau$.

Defining an averaged electric field squared for the gamma beam as $\langle E^2(x)\rangle=N/(A T)$ and the analogous quantity for the signal as $\langle E_\perp^2(x)\rangle=N_\perp/(A_{\rm s}\tau)$, we find
$\langle E_\perp^2(x)\rangle/\langle E^2(x)\rangle =(N_\perp/N)(T/\tau)(w_\gamma/w_0)^2$ and introduce the ellipticity angle of the beam as $\chi|_{\beta=\frac{\pi}{4}}=\arctan\bigl(\sqrt{\langle E_\perp^2(x)\rangle/\langle E^2(x)\rangle}\bigr)$; cf. also Eqs.~\eqref{eq:chi} and \eqref{eq:NperpbyN} above.
From these results we can estimate the total phase difference accumulated by the gamma beam in the parameter regime where $w_\gamma\gg w_0$ and $T\gg\{\tau,{\rm z}_{\rm R}\}$ as
\begin{equation}
 \Phi\simeq\frac{2^{7/4}\alpha^2}{15\sqrt{\pi}}\,\frac{W}{m_e}\,\frac{\omega}{m_e}\sin(2\beta)\Bigl(\frac{\lambdabar_{\rm C}}{w_0}\Bigr)^2\,\frac{8{\rm z}_{\rm R}}{\tau}\,{\rm e}^{\frac{1}{2}(\frac{8{\rm z}_{\rm R}}{\tau})^2}\,{\rm erfc}^{1/2}\bigl(\tfrac{8{\rm z}_{\rm R}}{\tau}\bigr)\simeq0.13\,.
 \label{eq:Philaser}
\end{equation}
The fairly large explicit value given here follows from an explicit evaluation of this expression for the high-intensity laser parameters given above, a gamma photon energy of $\omega\simeq400\,{\rm MeV}$ and an angle of $\beta=\pi/4$ between the polarization vectors of the pump and probe laser fields maximizing the signal.
A measurement of this value should be possible with pair-production polarimetry \cite{Eingorn:2018,Chattopadhyay:2021mbb}.
On the other hand, when plugging the explicit high-intensity laser parameters given above into Equation~\eqref{eq:Nperp} and utilizing the approximation~\eqref{eq:frac*F} with $\omega\simeq400\,{\rm MeV}$, $T\simeq160\,{\rm fs}$ and $w_\gamma\simeq20\,\upmu{\rm m}$, we obtain $N_\perp/N\simeq2.07\times10^{-6}$.

Note that as opposed to the result for the number of polarization-flipped signal photons in \Eqref{eq:Nperp}, the phase difference in \Eqref{eq:Philaser} does not feature an explicit dependence on the pulse duration $T$ and beam radius $w_\gamma$ of the gamma beam in the interaction area with the high-intensity laser beam. In fact, apart from the photon energy of the probe light, in the considered parameter regime the induced phase difference $\Phi$ is fully controlled by the parameters of the driving high-intensity laser field.

It is also instructive to note that upon trading the dependence on the pulse energy for the peak field amplitude by using \Eqref{eq:peakfield},  \Eqref{eq:Philaser} can be recast as
\begin{equation}
 \Phi\simeq {\rm z}_{\rm R} \omega\sin(2\beta)\frac{2^{1/4}\alpha}{30}\Bigl(\frac{e{\cal E}_0}{m_e^2}\Bigr)^2\,{\rm e}^{\frac{1}{2}(\frac{8{\rm z}_{\rm R}}{\tau})^2}\,{\rm erfc}^{1/2}\bigl(\tfrac{8{\rm z}_{\rm R}}{\tau}\bigr)\,,\label{eq:Philaserfield}
\end{equation}
which closely resembles the structure of \Eqref{eq:PhiBconst}. The phase difference scales quadratically with the field strength of the pump field (see Fig.~\ref{fig:feynman}), and linearly with the probe photon energy multiplied by the typical extent of the pump field along the propagation direction of the probe. In the former scenario this length is given by the length of the magnets $l$ and in the latter typically by the Rayleigh range ${\rm z}_{\rm R}$ of the high-intensity laser beam.

As a consistency check, we compare our result~\eqref{eq:Philaserfield} with the finding of Ref.~\cite{Heinzl:2006xc} utilizing a crossed-constant field calculation to analyze a high-intensity laser based vacuum birefringence scenario in the parameter regime where ${\rm z}_{\rm R}\ll\tau$. In this limit and for $\beta=\pi/4$, \Eqref{eq:Philaserfield} becomes  $\Phi\simeq{\rm z}_{\rm R} \omega(2^{1/4}\alpha/30)(e{\cal E}_0/m_e^2)^2$, while the corresponding result of Ref.~\cite{Heinzl:2006xc} is  $\Phi_{\text{\cite{Heinzl:2006xc}}}\simeq{\rm z}_{\rm R} \omega(2\alpha/15)(I_0/I_{\rm cr})$, with peak intensity $I_0\simeq2W/(\pi w_0^2\tau)= {\cal E}_0^2\sqrt{\pi/32} $ and critical intensity $I_{\rm cr}=(m_e^2/e)^2$; cf. also Ref.~\cite{Karbstein:2016lby}. Correspondingly, we find $\Phi/\Phi_{\text{\cite{Heinzl:2006xc}}}\simeq(2\sqrt{2}/\pi)^{1/2}\approx0.95$, which implies an excellent agreement.

In contrast to the case of a static magnetic field discussed above, for the scenario involving a high-intensity laser pulse only a fraction of the total number of gamma photons provided by the Gamma Factory is  available for testing the vacuum birefringence phenomenon: the repetition rate of the experiment is limited by the repetition rate of ${\cal O}(1)\,{\rm Hz}$ of the high-intensity laser.

\section{Conclusions}\label{sec:concls}

In this article we have briefly studied the perspectives of inducing a sizable QED vacuum birefringence signal at the Gamma Factory. After assessing in detail the parameter regime which can be reliably studied resorting to the leading contribution to the Heisenberg-Euler effective Lagrangian, we considered two specific scenarios giving rise to a vacuum birefringence phenomenon for traversing gamma probe photons.
In the first scenario the effect is driven by a macroscopic magnetic field provided by a set of LHC magnets, and in the second one by a counter-propagating high-intensity laser pulse.
For both cases, we explicitly determined the values of the phase difference characterizing the ellipticity acquired by the gamma probe as well as the associated numbers of polarization flipped signal photons.

It remains to be seen if the predicted signals are large enough to be accessible in dedicated experiments employing state-of-the-art technology.

\begin{acknowledgments}

I am indebted to Dmitry Budker, Elena Mosman and Alexey Petrenko for helpful discussions.
This  work  has  been  funded  by  the  Deutsche Forschungsgemeinschaft  (DFG)  under  Grant  No. 416607684 within the Research Unit FOR2783/1.

\end{acknowledgments}

\end{document}